\documentclass[10pt]{iopart}

\usepackage{pstricks,pst-node,pst-text,pst-3d}

\begin{document}

\title{Analog cosmology with spinor Bose-Einstein condensates}
\author{Esteban Calzetta\\
{\small CONICET and Departamento de Fisica, FCEN,}\\
{\small Universidad de Buenos Aires- Ciudad Universitaria, 1428 Buenos
Aires, Argentina}}
\date{\today }

\begin{abstract}
It is well known that both a particle with spin and a quantum field in a Bianchi type IX Universe are related to the quantum mechanical top. In this paper we will combine these insights to show that the dynamics of a cold gas of noninteracting atoms with total angular momentum $F\neq 0$ may be used to explore the behavior of quantum matter near a cosmological singularity.

\end{abstract}
\pacs{03.75.Kk, 04.62.+v, 03.70.+k}
\noindent{\it Keywords}: BEC, cosmology
\maketitle

\section{Introduction}
It is well known that both a particle with spin \cite{SCH68,SCH81} and a quantum field in a Bianchi type IX Universe \cite{Hu72,Hu73,Hu74} are related to the quantum mechanical top. In this paper we will combine these insights to show that the dynamics of a cold gas of noninteracting atoms with total angular momentum $F\neq 0$ may be used to explore the behavior of quantum matter near a cosmological singularity.

The physics of Bose-Einstein condensates (BEC) is one of the most active fields of reearch in Physics both in the experimental and theoretical fronts \cite{PS02}. Not one of the least reasons for this activity is that BEC provides the possibility of high accuracy tests of fundamental mechanisms of interest to other fields such as many-body \cite{BDZ07}, condensed matter \cite{LSADSS07}, high-energy physics \cite{RCPR07} and, most relevant to our present concerns, gravitation and cosmology \cite{BaliViLivRev,UnrSch07}.

There are two basic strategies by which BEC may be used to test fundamental issues in gravitation and cosmology. On one hand, it is well known that long wavelength excitations of a condensate propagate on an effective metric depending on the inhomogeneities in the condensate itself. If these may be controlled, then the effective matric may be designed to correspond to that of interest, for example, a Friedmann-Robertson-Walker \cite{JWVG07} or a Black Hole space time \cite{BaFaFa07,CFRBF08}.

One drawback of several of these standing proposals for using BEC as analog to gravitation and cosmological physics is that the focus is not on the behavior of the condensate itself, but of the quantum fluctuations above it. This makes the relevant effects small and hard to access experimentally. In the following we shall present a model of analog semiclassical cosmology which is free of this problem. The model involves a $F\neq 0$ noninteracting cold bosonic gas subject to an external magnetic field. We claim the behavior of the cold gas is analog to a quantized field in a Bianchi type IX space time.

In a different type of approach, one focuses not on duplicating a whole space-time, but rather an specific effect, such as superradiance \cite{TaTsVo07} or signature change \cite{WeWhVi07}. One important example is the behavior of quantum fluctuations on horizon crossing, which may be probed both in the so-called \textsl{Bose Nova} experiment (see \cite{JILA01,Cla03,CoThWi06,ChVoKe03} for the experiment, \cite{CH03,CH05,Cal08} for a theoretical interpretation) and in the superfluid - Mott insulator transition (again, see \cite{GMEHB02,SMSKE04,KMSSE05} for the experiment, and \cite{FiScUh07} for a discussion in the context of analog cosmology). As it may be seen, this second approach makes a much stronger connection with experiment.

The approach to analog cosmology which we shall present here shares properties of both approaches because, while it aims to duplicate a cosmological metric (in this case, a Bianchi type IX \textsl{Mixmaster} Universe) in BEC experiments, the experiments in question are easily within present means, and in some cases they have actually already been performed (see below).

The Bianchi type IX or Mixmaster model \cite{LLCF,MTW,RS75} is the homogeneous Universe whose metric has the largest number of free parameters and as such it has been proposed as a likely model for the Universe near the cosmological singularity \cite{BKL70}. It has a fascinating dynamics already at the classical level \cite{Bar82,HBC94}. It has also been investigated at the quantum \cite{Mis70,Mis72} and semiclassical \cite{Hu72,Hu73,Hu74,Cal91} levels.

\section{Spinor BEC and Bianchi type IX Universes}
Spinor BEC occur in species where the interaction of the nuclear spin $\mathbf{I}$ and the electron spin $\bf{S}$ for an electron in the ground state introduces in the Hamiltonian a hyperfine term $(A/2)\bf{F^2}$, where $\bf{F=I+S}$ and $A$ is a constant (for rubidium, $A\approx 2\;10^{-6}$eV), leading to a ground state with ${\bf F}^2=F(F+1)\neq 0$ \cite{Ho98,OhmMac98,PS02}. Spinor BEC have been realized experimentally with several different species: sodium $^{23}$Na with $F=1$ \cite{SISMCK98,BGTJL07} and $F=2$ \cite{GGLLCGIPK03}; rubidium $^{87}$Rb with $F=1$ \cite{CHBSFZYC04,CQZYC05,KBBWNBS05,SHLVS06,ALBSPP07,VLGS07} and $F=2$ \cite{SEKKSCABS04,KAEH04,KBNBS06}  and chromium $^{52}$Cr with $F=3$ \cite{LKFFMGGP07,BCZLMVKG07,LMFKM08}. Spinor BEC can also be achieved with Rydberg atoms and molecular condensates \cite{MLLP07}. See also \cite{Lam07,UhScFi07,DamZur07a,DamZur07b,SaKaUe07a,SaKaUe07b,MiaGir07,CGSD07,SaiHyu08}. In general the condensate atoms will interact among themselves. However, we shall assume these interactions can be made negligible, for example, by exploiting the dependence of the scattering length with respect to external fields \cite{KKHV02,MVSDRKV02}, or simply by working with a dilute enough gas. See \cite{VLGS07}, where the dipolar interaction is suppressed by forcing the dipoles to rotate, and \cite{SISMCK98,LKFFMGGP07}, where the scattering lengths for the local and spin exchange interactions are manipulated in an actual experimental setting.

There are several ways to develop a path integral quantization for a spinor BEC, besides simply associating a different field to each angular momentrum projection. For example, we may use spin coherent \cite{Kla79,ShiTak99,StPaGa00,Gri03,Gri06} or modified spin coherent states \cite{AltAue02,PADHL05}. In the formalism advanced by Schulman \cite{SCH68,SCH81} spin is described by adscribing to the spinning particle an internal space parameterized by three Euler angles $\theta$, $\varphi$ and $\psi$. Introducing the Euler parameters \cite{Ros51} 


\begin{eqnarray}
X&=&\cos\:\frac{\theta}{2}\;\cos\:\frac{\varphi +\psi}{2}\\
Y&=&\cos\:\frac{\theta}{2}\;\sin\:\frac{\varphi +\psi}{2}\\
Z&=&\sin\:\frac{\theta}{2}\;\cos\:\frac{\varphi -\psi}{2}\\
W&=&\sin\:\frac{\theta}{2}\;\sin\:\frac{\varphi -\psi}{2}
\label{Euparam}
\end{eqnarray}
it is clear that the internal space has the topology of the three dimensional sphere $S^3$ embedded into a four dimensional space. The usual euclidean metric induces a metric on the sphere

\begin{equation}
ds_0^2=d\theta^2+d\varphi^2+d\psi^2+2\cos\theta\:d\varphi d\psi
\label{0metric}
\end{equation}
and a natural volume element $d^3\chi\;\sqrt{g_0}$, where the coordinates $\chi^a$, $a=1$, $2$, $3$ are the three angles $\theta$, $\varphi$ and $\psi$, respectively and $g_0=\sin^2\theta$ is the determinant of the metric.

Upon second quantization we describe the cold gas by a q-number field $\Psi\left(\bf{x},\theta,\varphi,\psi\right)$ defined on $R^3\times S^3$. Under the one-mode approximation, the external and internal dependences decouple, and we shall discuss only the later. We note that, if required, a position-independent condensate may be achieved in practice \cite{MSHCH05}. 

As in the case of a quantum rotor, we identify the components $F_i$ of angular momentum with the operators

\begin{eqnarray}
F_x&=&\frac \hbar i \left\{\cos\left[\varphi\right]\frac{\partial}{\partial\theta}-\frac{\sin\left[\varphi\right]\cos\left[\theta\right]}{\sin\left[\theta\right]}\frac{\partial}{\partial\varphi}+\frac{\sin\left[\varphi\right]}{\sin\left[\theta\right]}\frac{\partial}{\partial\psi}\right\}\\
F_y&=&\frac \hbar i \left\{\sin\left[\varphi\right]\frac{\partial}{\partial\theta}+\frac{\cos\left[\varphi\right]\cos\left[\theta\right]}{\sin\left[\theta\right]}\frac{\partial}{\partial\varphi}-\frac{\cos\left[\varphi\right]}{\sin\left[\theta\right]}\frac{\partial}{\partial\psi}\right\}\\
F_z&=&\frac \hbar i\frac{\partial}{\partial\varphi}
\end{eqnarray}
They satisfy the usual commutation relations

\begin{equation}
\left[F_i,F_j\right]=i\hbar\epsilon_{ijk}F_k
\end{equation}
We note that there are also the projections $F_a$ of angular momentum on the axes of inertia $\xi$, $\eta$ and $\zeta$ of the quantum top

\begin{eqnarray}
F_{\xi}&=&\frac \hbar i\frac{\partial}{\partial\psi}\\ 
F_{\eta}&=&\frac \hbar i 
\left\{\cos\left[\psi\right]\frac{\partial}{\partial\theta}-\frac{\sin\left[\psi\right]\cos\left[\theta\right]}{\sin\left[\theta\right]}\frac{\partial}{\partial\psi}+\frac{\sin\left[\psi\right]}{\sin\left[\theta\right]}\frac{\partial}{\partial\varphi}\right\}\\
F_{\zeta}&=&\frac \hbar i\left\{-\sin\left[\psi\right]\frac{\partial}{\partial\theta}-\frac{\cos\left[\psi\right]\cos\left[\theta\right]}{\sin\left[\theta\right]}\frac{\partial}{\partial\psi}+\frac{\cos\left[\psi\right]}{\sin\left[\theta\right]}\frac{\partial}{\partial\varphi}\right\}
\end{eqnarray}
whose commutation relations are

\begin{equation}
\left[F_a,F_b\right]=-i\hbar\epsilon_{abc}F_c
\end{equation}
The Hamiltonian for the second quantized free spinor gas is

\begin{equation}
H_0=\int d^3\chi\;\sqrt{g_0}\;h_0
\end{equation}
with a Hamiltonian density, in natural units $\hbar=1$
\begin{equation}
h_0=\frac{A}{2}\:g_0^{ab}\frac{\partial\Psi^{\dagger}}{\partial\chi^a}\frac{\partial\Psi}{\partial\chi^b}
\label{hache0}
\end{equation}
 $g_0^{ab}$ is the inverse metric  on the sphere.

The sphere is also a spatial section of the so-called Einstein Universe. BEC in the Einstein Universe has been discussed in \cite{Alt78,AltMal00,Alt02,ParZha91}. See also \cite{Tom92,KirTom95,SmiTom96}.

If we now turn on an external magnetic field $B$ in the $z$ direction, there will be new terms in the Hamiltonian proportional to the magnetic moments of electron and nucleus. Since the latter is negligible compared to the former, only the electron magnetic moment needs to be considered. The new term couples $F=1$ states among themselves, but also to $F=2$ states. If we integrate out these more energetic states, we obtain an effective Hamiltonian density for the $F=1$ states which involves a linear and a quadratic term in the external field. We may also consider the quadratic Zeeman effect which comes from terms which are quadratic on the external field in the microscopic Hamiltonian \cite{SchSny39,ClaTay82} The linear term may be minimized by a suitable choice of initial condition \cite{SHLVS06}, or else eliminated by transforming into a rotating frame, and we shall not consider it. The quadratic term reads \cite{SaKaUe07a}

\begin{equation}
\frac{\mu_B^2B^2}{cA}\int d^3\chi\;\sqrt{g_0}\frac{\partial\Psi^{\dagger}}{\partial\varphi}\frac{\partial\Psi}{\partial\varphi}
\end{equation}
where $\mu_B$ is the Bohr magneton, and $c$ is a constant. For $F=1$ rubidium, $c=8$.  This has the effect of changing the metric of the internal space into 

\begin{equation}
ds_B^2=\left[1+{\cal B}^2\right]d\theta^2+d\varphi^2+\left[1+{\cal B}^2\sin^2\theta\right]d\psi^2+2\cos\theta\:d\varphi d\psi
\label{Bmetric}
\end{equation}
where

\begin{equation}
{\cal B}=\sqrt{\frac 2c}\frac{\mu_BB}{A}
\label{calB}
\end{equation}
To see that this indeed describes a Bianchi type IX Universe, we introduce three nonintegrable differential forms $\omega_a$

\begin{eqnarray}
\omega_1&=&d\varphi +\cos\theta\;d\psi\\
\omega_2&=&\cos\varphi\,d\theta +\sin\varphi\sin\theta\;d\psi\\
\omega_3&=&-\sin\varphi\,d\theta +\cos\varphi\sin\theta\;d\psi
\label{forms}
\end{eqnarray}
These forms satisfy
\begin{equation}
d\omega_a=-\frac{1}{2}\epsilon_{abc}\omega_b\wedge\omega_c
\label{diffs}
\end{equation}
where $\epsilon_{abc}$ is the totally antisymmetric symbol. The metric can then be written in terms of Misner parameters \cite{Mis70} as

\begin{equation}
ds_B^2=e^{2\Omega}\left\{e^{-2\beta_+}\omega_1^2+e^{\beta_++\sqrt{3}\beta_-}\omega_2^2+e^{\beta_+-\sqrt{3}\beta_-}\omega_3^2\right\}
\label{Charlie}
\end{equation}
where $\beta_-=0$, $\beta_+=\Omega$ and

\begin{equation}
\Omega =\frac{1}{3}\ln\left[1+{\cal B}^2\right]
\label{Omega}
\end{equation}

When $\beta_-=0$ the Bianchi type IX Universe reduces to the so-called Taub model \cite{RS75}. BEC in the Taub Universe, in the limit of small anisotropy, has been discussed in \cite{Hua94}.

This concludes our discussion. Observe that the Bianchi type IX metric appears as the metric in internal space, not as an effective metric describing the evolution of long wavelength fluctuations above the condensate. Therefore any nontrivial effect due to geometry will be much more readily apparent in this problem than in other competing schemes. Moreover, the Bianchi type IX Mixmaster Universes are one of the most fascinating models in cosmology. We expect this short note will open the door to new crossfertilization between BEC and gravity.

\section*{Acknowlegments}
This paper is dedicated to Bei-lok Hu on his 60th birthday.

I thank H. Grinberg and S. Liberati for discussions.

This work was completed during a stay at the Kavli Institute for Theoretical Physics (KITP), University of California - Santa Barbara. A preliminary version was presented at the meeting on Nonequilibrium phenomena in cosmology and particle physics held at KITP, Feb. 2008

This work is supported in part by ANPCyT, CONICET and UBA, Argentina.

\end{document}